\documentclass[prb,twocolumn,showpacs,superscriptaddress,floatfix]{revtex4}

\usepackage[pdftex]{graphicx}
\usepackage[pdftex]{color}
\begin{document}
\title{Critical state analysis of orthogonal flux interactions in pinned superconductors}
\author{A. Bad\'{\i}a\,--\,Maj\'os}
\email[Electronic address: ]{anabadia@unizar.es}
\affiliation{Departamento de F\'{\i}sica de la Materia Condensada--I.C.M.A., Universidad de Zaragoza--C.S.I.C., Mar\'{\i}a de Luna 1, E-50018 Zaragoza, Spain}
\author{C. L\'opez}
\affiliation{Departamento de Matem\'aticas, Universidad de Alcal\'a de Henares, E-28871 Alcal\'a de Henares, Spain}

\date{\today}
%
%\vspace{-5mm}
%
\begin{abstract}

We show that, based on the critical state model for flux-line pinning in hard superconductors, one can assess the magnetic moment relaxation induced by the oscillations of a perpendicular magnetic field. Our theory follows a recent proposal of using phenomenological 2D modeling for the description of crossed field dynamics in high-T$_c$ superconductors [{\tt arXiv:cond-mat/0703330}]. 

Stationary regimes with either saturation to metastable configurations, or complete decay to the thermodynamic equilibrium are obtained. The transition between both types of response is related to the disappearance of a flux free core within the sample. As a common feature, a step-like dependence in the time relaxation is predicted for both cases. The theory may be applied to long bars of arbitrary and non homogeneous cross section, under in-plane magnetic field processes.

\end{abstract}
%
%\vspace{10mm}
%
\maketitle

\section{Introduction}
As it was emphasized by Brandt and Mikitik,\cite{PRLshake} magnetic relaxation in superconductors is not always related to thermally activated flux creep. Thus, one can predict the magnetic moment relaxation induced by perpendicular ac field oscillations, within the critical state framework. Recall that such phenomenological approach\cite{bean} describes the zero temperature limit of magnetic flux dynamics in superconductors with pinning. The basic idea is that a given magnetic source produces some distribution of persistent current densities (critical currents) within the sample, determined by the actual excitation process through Faraday's law.

Essentially, the model in Ref.\onlinecite{PRLshake} ascribes the magnetic moment relaxation to internal variations of the current density distribution, related to the shaking effect from the transverse ac field. It is of note that, within such a theory, and depending on the actual conditions of the shaking process, one can either predict relaxation toward the true equilibrium state of the superconductor, or eventual freezing into some metastable configuration. This feature seems to be tightly linked to the critical state physics. However, some recent research about crossed field effects in high-T$_c$ superconductors\cite{vanderbemden} has raised doubts about the generality of the complete/incomplete relaxation issues.

Dimensionality is a relevant property in the transverse flux phenomenon, because, at least, two component magnetic systems are involved. Then, as it was remarked by Brandt and Mikitik, the appropriate critical state problem to be considered is, minimally, two dimensional. In their work for thin samples, taking advantage of the smallness of the $thickness/width$ ratio a quasi-one-dimensional analysis was given. Namely, they split the problem into two coupled one-dimensional flux pinning statements. 

On the other hand, at this point, the reader may recall that critical state models for one-dimensional geometries (infinite slabs and cylinders) incorporating transverse flux effects have been issued in different forms.\cite{clem,badiaprb} However, the latter studies refer to magnetic field vectors with mutually perpendicular components, but parallel to the surface of the sample. This mathematical simplification implies an important physical assumption: the dynamics of transverse flux lines includes cutting effects between them. Clem\cite{clem} has justified that, still, the phenomenon may be treated within the critical state framework. This is done via the introduction of the so-called parallel (to $\vec{B}$) critical current density $J_{c\parallel}$. In contrast to the most familiar perpendicular critical current density $J_{c\bot}$, which relates to the flux pinning effect,\cite{bean} $J_{c\parallel}$ relates to the maximum angle between flux lines before undergoing a cutting process. It has been shown that a number of experimental observations can be understood within the flux pinning/cutting framework. These include the so-called {\em butterfly} and {\em collapse} effects in the transverse magnetic moment.\cite{clem,badiaprb}
\begin{figure}[b]
\centerline{
\includegraphics[width=7.5cm,angle=0]{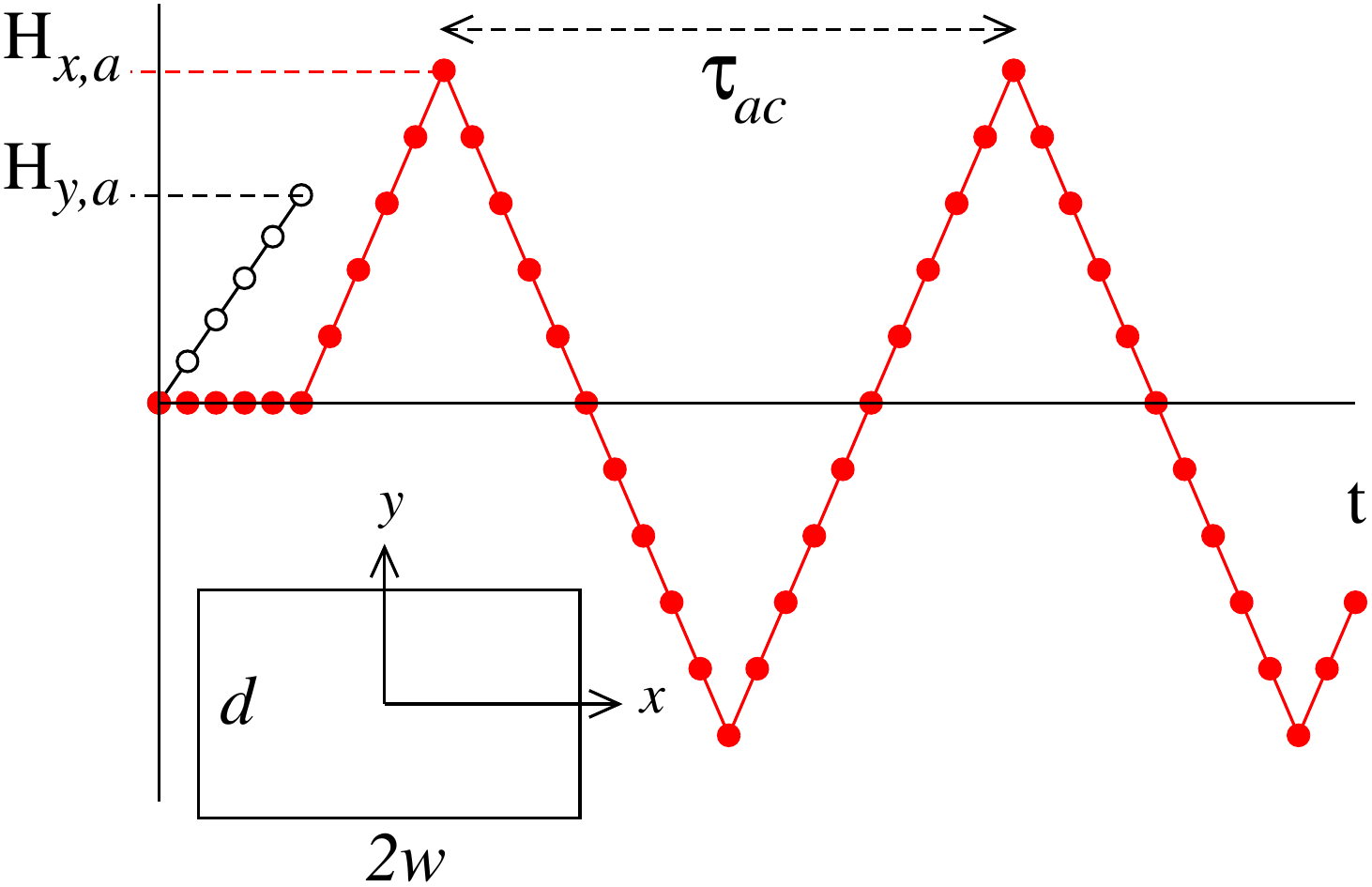}}
\caption{\label{fig1}Sketch of the experiment simulated in this work. An infinite bar of cross section $2w\times d$ is subjected to the magnetic field excitation shown in the plot. $H_y$ is initially increased to the value $H_{y,a}$ and then $H_x$ is cycled with periodicity $\tau_{ac}$ and amplitude $H_{x,a}$.}
\end{figure}

As a breakthrough in the above scenario, Vanderbemden {\em et al.}\cite{vanderbemden} have emphasized that, with moderate numerical effort, two-dimensional theories can be used for describing transverse flux effects. This rules out the necessity of considering either very thin samples or incorporating cutting phenomena. The modeling is realized by analyzing the evolution of in-plane magnetic fields within the cross section of a superconducting bar. In fact, it is apparent that the mutually perpendicular flux lines penetrate basically along (and not across) each other. Within this picture, one can analyze transverse flux effects only related to the flux pinning mechanism (the $J_{c\bot}$ limitation). However, the former investigations were done in a flux creep framework (power law $E\propto(J/J_{c})^{n}$), and for quite specific initial conditions. As a consequence, the authors did always observe complete relaxation by transverse shaking. In this work, we have studied the true critical state limit ($n\to\infty$ in the above $E-J$ law) for two-dimensional transverse flux problems under a wide range of conditions. By using this {\em ansatz}, it will be shown that either complete or incomplete flux relaxation may be expected.

The article is organized as follows. First, in Sec.\ref{secCritical}, we describe the in-plane magnetic flux configurations under study and justify the application of a generalized Bean's model for the investigation of such phenomena. A numerical method, based on the concept of mutual inductance between circuits will be introduced. In Sec.\ref{secCrossed}, a set of numerical results of the actual magnetic flux dynamics in a number of crossed field configurations is presented. Based on the critical current distributions, and also related to the structure of the penetrating field lines, we show the importance of the partial penetration concept in this problem. In close analogy to the {\em one dimensional} critical state problems, increasing the applied magnetic field progressively leads to the disappearance of the flux free region. Sec.\ref{secMagnetic} is devoted to analyze the relaxation of the magnetic moment, induced by oscillations of transverse magnetic fields. A systematic study for different amplitudes of the {\em polarizing} dc field, as well as of the {\em oscillating} transverse component is presented. Decay to the true equilibrium magnetization is clearly related to the disappearance of the flux free core. The generality of the above properties is shown by application of the model to samples with inhomogeneities in the cross section. Thus, in Sec.\ref{secExtensions} we illustrate the evolution of the magnetic flux structure for samples with non homogeneous critical current density $J_c$. 

\section{Critical State Model}
\label{secCritical}

The basic configuration considered in the present work is sketched in Fig.\ref{fig1}. A superconducting bar occupies the region defined by $|x|\leq w$, $|y|\leq d/2$, $|z|\leq\infty$. The external magnetic field will be applied along the $x$ and $y$ axes. Starting from a zero field configuration, $H_y$ will be ramped to a final amplitude $H_{y,a}$. Then $H_x$ will be cycled between the values $\pm H_{x,a}$ in a linear fashion. We note that, corresponding to the hypothesis that the lower critical field of the superconductor may be neglected as compared to $H_{x,a}$ and $H_{y,a}$, the equality $\vec{B}=\mu_{0}\vec{H}$ will be used in what follows. In other words, the equilibrium magnetization of the sample is approximated by zero.

The starting point for the application of the critical state theory to the above problem is our variational statement\cite{badiaprb} within the $\{\vec{A},\vec{J}\,\}$-formulation.\cite{badiapl} Such a representation in terms of the vector potential and current density allows to include finite size effects in a simple manner, and has been well elaborated in previous work. The equivalence to the more standard differential equation statements based on the Maxwell equations is rigorously justified in Ref.\onlinecite{badiajpa}. Thus, the quasi-stationary evolution of magnetic processes in a hard superconductor may be obtained from the constrained minimization of the functional
\begin{eqnarray}
\label{eqnJJ}
&{\cal F}\equiv{\displaystyle\int}_{\!\!\Omega}{\displaystyle\int}_{\!\!\Omega}
\left[
\displaystyle\frac{\vec{J}_{\rm n+1}(\vec{x})\cdot\vec{J}_{\rm n+1}\,(\vec{x}\,')}{|\vec{x}-\vec{x}\,'|}
-2\frac{\vec{J}_{\rm n}(\vec{x})\cdot\vec{J}_{\rm n+1}\,(\vec{x}\,')}{|\vec{x}-\vec{x}\,'|}
\right]&
\nonumber\\
&+{\displaystyle\frac{8\pi}{\mu_{0}}\int}_{\!\!\Omega}
\left(
\vec{A}_{\rm e,n+1}-\vec{A}_{\rm e,n}
\right)\cdot\vec{J}_{\rm n+1}(\vec{x})\; .& 
\end{eqnarray}
Here, $\vec{J}_{\rm n}$ stands for the current density at the time layer $n\delta t$, $\Omega$ represents the superconducting region, and $\vec{A}_{\rm e}$ means the externally applied vector potential. Minimization is performed iteratively in time with $\vec{J}_{\rm n+1}\,(\vec{x})$ the unknown function for each step. As regards the constraints, recall that, for the present work, the minimization of ${\cal F}$ under the condition $\|\vec{J}\|\leq J_{c}$ will be equivalent to the flux pinning criterion $|J_{\bot}|\leq J_{c\bot}$. Indeed, by symmetry arguments one can show that $\vec{J}$ strictly flows along the bar, i.e.: $\vec{J}=(0,0,J_{z})$ and then $\|\vec{J}\|=|J_{\bot}|$.

On the technical side, an important advantage of the above mentioned problem's symmetry is the simplification of the numerical algorithm to be used for solving the variational statement. Thus, one can argue that the current density streamlines may be replaced by a collection of infinite straight wires, and the variational statement can be easily discretized to an algebraic form. The problem is transformed into the minimization of the quadratic function
\begin{eqnarray}
{\tt F}\equiv &&{\displaystyle \frac{1}{2}}\sum_{i,j}I_{\rm i,n+1}M_{ij}I_{\rm j,n+1}
-\sum_{i,j}I_{\rm i,n}M_{ij}I_{\rm j,n+1}
\nonumber\\
&&+\mu_{0}\sum_{i}I_{\rm i,n+1}(A_{\rm e,n+1}-A_{\rm e,n}) \, .
\label{eqMij}
\end{eqnarray}
Here, the set of unknowns $\{ I_{i}\}$ represents the collection of current lines flowing across the section of the sample, and $M_{ij}$ is their mutual inductance matrix. Minimization will be made under the set of constraints $-I_{c}\leq I_{i}\leq I_{c}$. The parameter $I_{c}$ is the critical current for an elementary wire, related to the flux pinning current density limitation. The mutual inductance coefficients will be evaluated from the expressions
\begin{eqnarray}
\label{eqninductances}
M_{ii}&=& \frac{\mu_0}{8\pi}
\nonumber\\
M_{ij}&=& \frac{\mu_0}{2\pi} {\rm ln}\frac{a^2}{(x_{i}-x_{j})^2+(y_{i}-y_{j})^2} \, ,
\end{eqnarray}
which one can be obtain for a collection of parallel straight wires of circular cross section (radius $a$).\cite{badiajp}

In summary, the calculation method is as follows. First, we compute the matrix $M_{ij}$ for a given grid of elements, describing the problem (most of the examples presented later correspond to the rectangular case of $\ell\times m =46\times 92$ elements). Second, we solve the minimization of the function in Eq.(\ref{eqMij}). Recall that one has $\ell\times m$ variables ($\{ I_{i}\}$) constrained by $\ell\times m$ conditions. This process is done iteratively in time ($n=0,1,2,\dots$) and one introduces the desired external bias, through the position dependent vector potential. In the case of a uniform field with components $(H_{x},H_{y},0)$ one can use the $z-$component $A_{e}=H_{x}y-H_{y}x$. 

Finally, we want to mention some additional details that may be of help from the practical point of view. On the one side, the further use of symmetry considerations related to the homogeneity of the applied magnetic field is to be advised. Thus, if one considers the rectangular cross section in the inset of Fig.\ref{fig1}, the inversion property $I(x,y)=-I(-x,-y)$ may be recalled. Only one half of the variables have to be accounted for, and the computation simplifies noticeably. On the other hand, an important advantage of the preferred $\{\vec{A},\vec{J}\,\}$-formulation is that the physical quantities of interest are obtained by integration of the former. Then, discretization errors are smoothed. On the contrary, if one uses an $\{\vec{H}\}$-formulation, $\vec{J}$ is obtained by differentiation and this results in the magnification of errors.

\section{Crossed field dynamics}
\label{secCrossed}
Below, we will present the calculated current density distributions and the corresponding magnetic field configurations induced by external excitations in the form illustrated in Fig.\ref{fig1}. For further analysis, we have performed calculations for a variety of values of the parameters $H_{x,a}$ and $H_{y,a}$. Both have been allowed to range well above and below the characteristic field $H_{p}$, which determines the full penetration regime for $\vec{H}$ along the $x$ axis. Within the geometry considered in this work ($w=d$) one has\cite{brandtstrip} $H_{p}\simeq 0.85 w J_{c}/2$.

Hereafter, dimensionless units for the magnetic field will be used, in terms of $H_{p}$, i.e.: $h\equiv H/H_{p}$.
\begin{figure}[t]
\centerline{
\includegraphics[width=8.5cm]{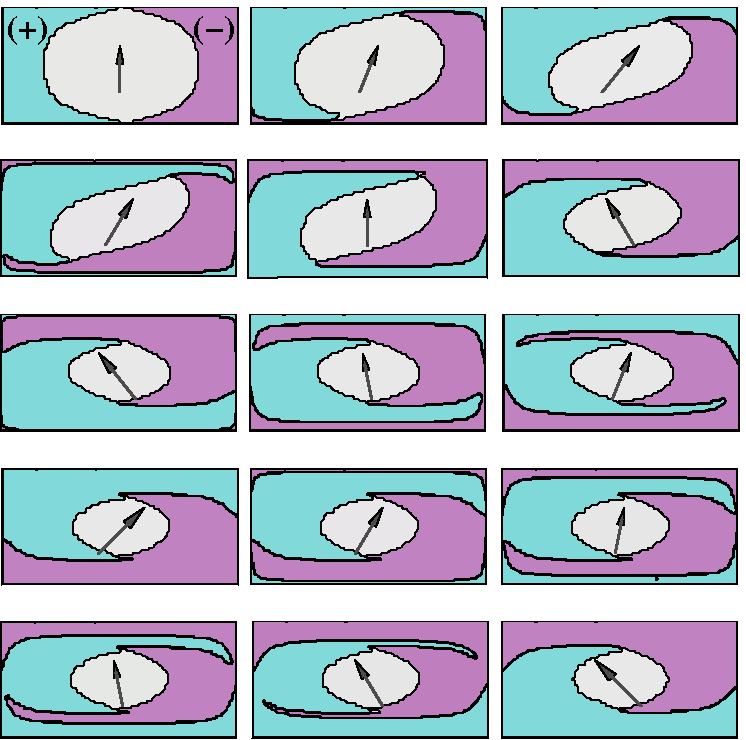}}
\caption{\label{fig2}(Color online) Critical current distributions within the cross section of the superconducting bar, subjected to the process in Fig.\ref{fig1}. The applied magnetic field vector is indicated by the central arrow. (+) indicates outward current flow ($J_c$), while (-) indicates inward current flow ($-J_c$). The central core is the subcritical region $J = 0$. The normalized values $h_{x,a}=h_{y,a}=0.4$ have been used for this plot.}
\end{figure}
\subsection{Critical currents}
As one expects from theoretical considerations,\cite{badiaprb} the numerical solution of Eq.(\ref{eqMij})
produces a current density distribution with the unknowns $I_{i}$ taking the values $\pm I_{c},0$. The actual structure of critical/subcritical zones within the bar's cross section strongly depends on the $\{ H_{x}(t),H_{y}(t)\}$ process. Below, we describe some systematic properties that have been observed.

\subsubsection{Partial penetration}

In Fig.\ref{fig2} we depict the critical current distributions in the initial part of the process illustrated in Fig.\ref{fig1}. To the upper left one can find the typical penetration profile for a uniform external field along one of the symmetry axes of the sample ($Y$-axis). In this case, this has been achieved when the condition $h_{y,a}=0.4$ is met. The subsequent pictures correspond to a number of intermediate steps within the first cycle for the $h_{x}$ component. This has been also ramped to the amplitude $h_{x,a}=0.4$. The actual orientation of the applied field is indicated by the central vector, that is proportional to $(h_{x},h_{y})$ within each picture. Notice the similarity between the current free cores within this study and the analytical results in Ref.\onlinecite{mikitik} for thin samples in oblique fields.

Two facts are noticeable in the evolution of the current density penetration profiles: (i) the introduction of $h_{x}$ induces upper and lower sheets of current, devoted to the shielding of the magnetic field variations, (ii) within the chosen range of parameters, one gets a central current free core. The core is initially distorted by the application of $h_{x}$, but rapidly acquires a stationary behavior. This property is apparent in the lower shots of Fig.\ref{fig2}, where we display the evolution within the first half of the second cycle in $h_{x}$. Recall that one obtains a dynamical structure for the $\pm I_{c}$ regions, around a stable central core with $I_{c}=0$.

\subsubsection{Full penetration}
\label{seccurrfull}

\begin{figure}[t]
\centerline{
\includegraphics[width=8.5cm]{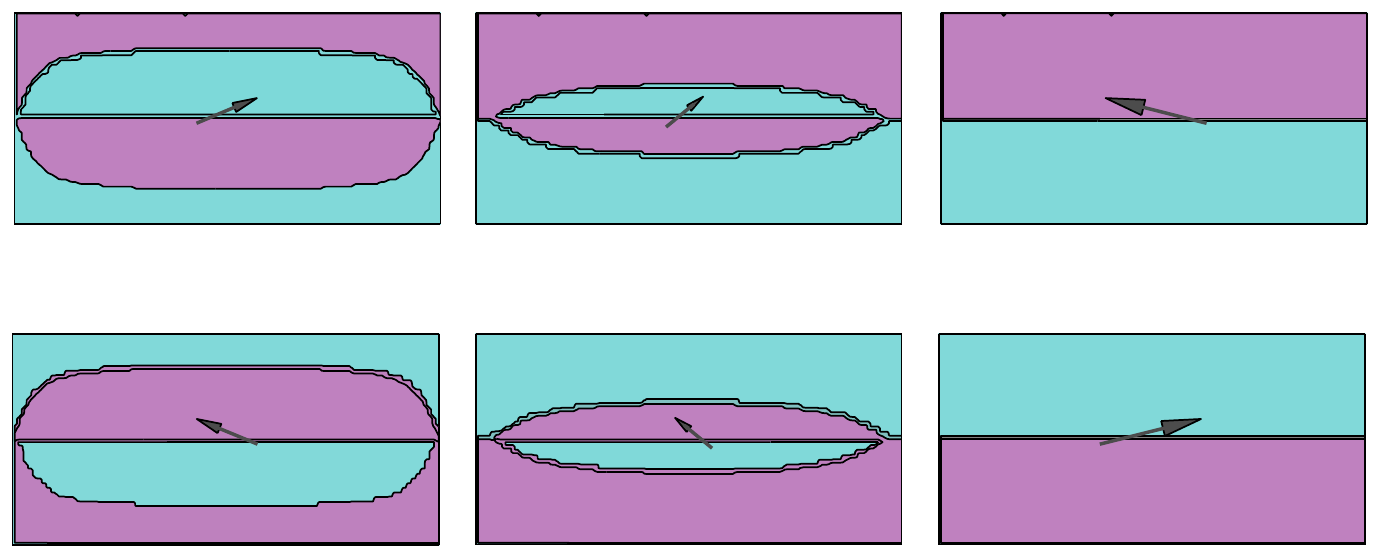}}
\caption{\label{fig3}(Color online) Same as Fig.\ref{fig2}, but using the values $h_{x,a}=1.6$ and  $h_{y,a}=0.4$. Only the stationary regime after several cycles is shown.}
\end{figure}

As one could infer from the knowledge of one-dimensional critical state problems, the subcritical region collapses to zero volume when the applied magnetic field reaches a certain intensity. In the two-dimensional case this is also true, but one can obtain a {\em full penetration} regime by combining $\{ h_{x,a},h_{y,a}\}$ and the actual excitation process in a number of ways. In Fig.\ref{fig3} we show the current density distributions obtained for the stationary regime in the case $h_{x,a}=1.6,h_{y,a}=0.4$. Recall that, together with the disappearance of the core, any trace of the initial critical state, related to the application of $h_{y,a}$ is lost. The stationary critical current distribution is only related to the oscillating $h_{x}$ component.

\subsection{Magnetic flux configurations}

\begin{figure}[b]
\centerline{
\includegraphics[width=8.5cm]{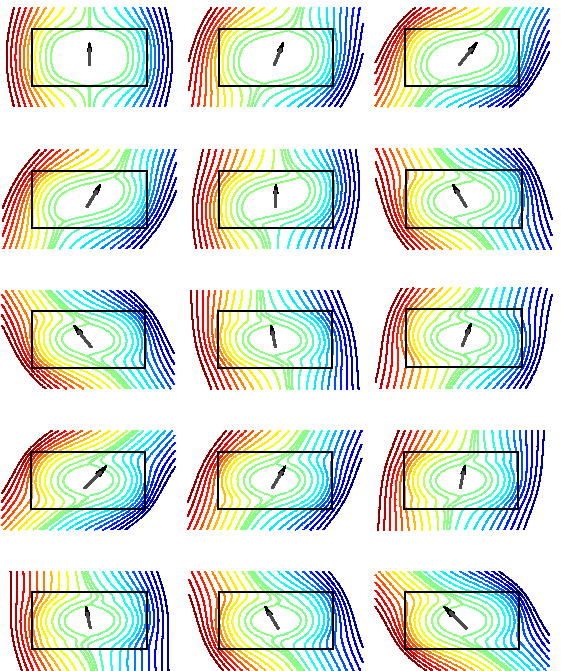}}
\caption{\label{fig4}(Color online) Magnetic flux lines for the process in Fig.\ref{fig2}. The flux free core is defined by the two lines with the lowest magnetic field modulus within our numerical resolution.}
\end{figure}

The resultant magnetic field profiles (applied plus induced) corresponding to the previous current density distributions are shown in Figs.\ref{fig4}--\ref{fig6}. The magnetic field lines have been obtained as the contour plots of the function $A_{z}(x,y)$, i.e.: they are given by the condition $A_{z}=constant$. This is a straightforward property in the 2D geometry of this work. On the other hand, the expression
\begin{equation}
A_i=\frac{\mu_{0}I_{i}}{4\pi}\ln \left[ \sqrt{(x-x_i)^{2}+(y-y_i)^{2}}/a \right]
\end{equation}
has been used in order to evaluate the vector potential contributed by an infinite wire of radius $a$, and carrying a current $I_{i}$.

\subsubsection{Partial penetration}

Fig.\ref{fig4} displays the magnetic field structure corresponding to the transient period in which the central core is established (upper rows). Recall that the magnetic field is excluded from the current free region (compare to Fig.\ref{fig2}) and also the trend of forming a stationary profile, surrounded by regions in which significant variations of the flux structure take place. Thus, in the lower rows, one can already observe that the inner core is unperturbed, while the applied field oscillates in between $\pm h_{x,a}$.

Fig.\ref{fig4} has been obtained by integration of the current density profiles presented in the previous section, and thus corresponds to the crossed field amplitudes $h_{y,a}=0.4$ and $h_{x,a}=0.4$.

\subsubsection{The flux free core}

\begin{figure}[t]
\centerline{
\includegraphics[width=6.5cm]{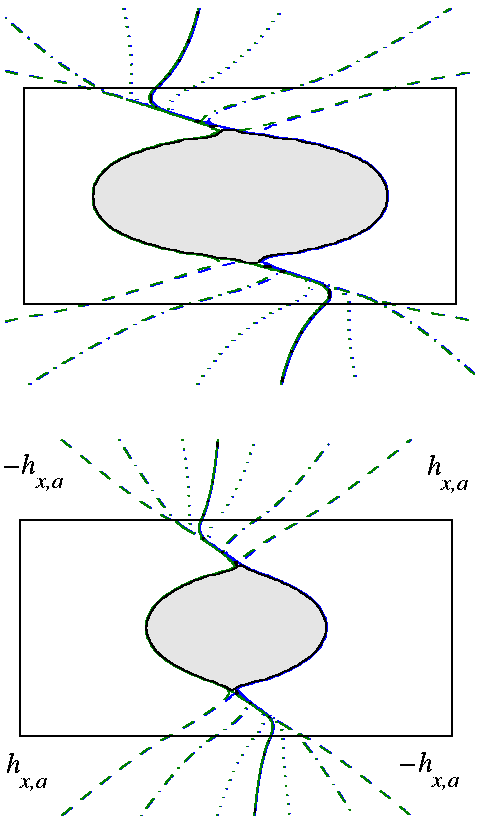}}
\caption{\label{fig5}Evolution of the magnetic field lines that define the stationary flux free core along the $\mp h_{x,a}$ cycle. The upper plot corresponds to the vertical amplitude $h_{y,a}=0.2$, while the lower one was obtained for $h_{y,a}=0.4$. Dashed, dot-dashed, dotted and continuous styles are used for the consecutive central line pair during the cycle.}
\end{figure}

In order to have a more systematic information about the partial penetration regime in 2D problems with crossed magnetic fields, we have studied the formation of the flux free core in different conditions. Thus, Fig.\ref{fig5} displays the stationary core that is obtained when one uses an oscillating field amplitude $h_{x,a}=0.4$, for initial ramps with either $h_{y,a}=0.2$ or $h_{y,a}=0.4$ respectively. The behavior for other values has been also studied, and the results (not shown for brevity) do not display significant differences. 

For the sake of clarity, in the plots, we have only shown the lowest level magnetic field lines that define the core, and their evolution as $h_{x}$ is cycled within the stationary regime. Notice that the boundary of the core remains unchanged, while the flux lines adapt to the applied field as one moves away from the center of the sample. The only difference between the two situations is the actual size of the core. As one can deduce from the results in the following section, this is straightforwardly related to the saturation value $m_y$ acquired by the vertical magnetic moment.

\subsubsection{Full penetration}

\begin{figure}[b]
\centerline{
\includegraphics[width=8.5cm]{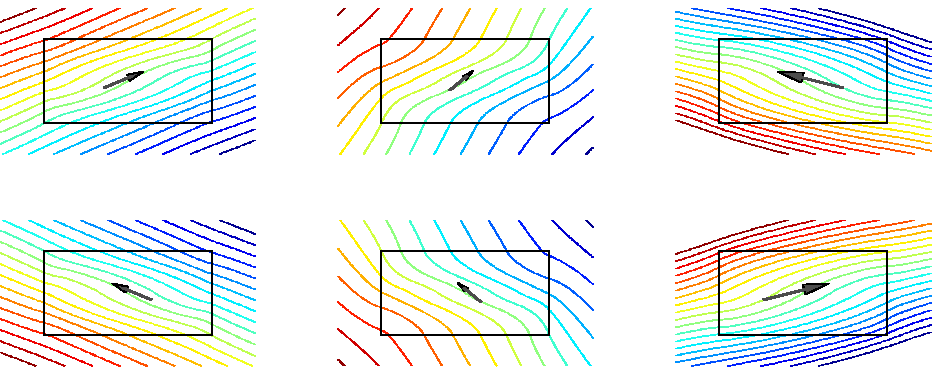}}
\caption{\label{fig6}(Color online) Magnetic flux lines corresponding to the full penetration profiles depicted in Fig.\ref{fig3}. As before, the central vector is proportional to the applied magnetic field.}
\end{figure}

As it was shown above (Sec.\ref{seccurrfull}), the current free core disappears when one uses the external conditions $h_{y,a}=0.4$ and $h_{x,a}=1.6$. Here we show the flux line structure corresponding to the oscillations of $h_{x}$ in such a range. As one can see in Fig.\ref{fig6}, magnetic flux has fully penetrated the sample. However, the over damped character of flux motion in the critical state is evident in that plot. Although the flux lines close to the periphery of the sample tend to follow the external excitation, the inner regions always presents a delay.

\section{Magnetic moment relaxation}
\label{secMagnetic}
The connection between the magnetic field profiles obtained in the previous section, and direct experimental observations is established below. As a figure of merit we have chosen the sample's magnetic moment (per unit length) both along the $x$ and $y$ axes. The definitions
\begin{eqnarray}
\label{eqm}
m_{x}&=&\int_{-w}^{w}dx \int_{-d/2}^{d/2}dy\,\left[yJ_{z}(x,y)\right]
\nonumber\\
\nonumber\\
m_{y}&=&\int_{-w}^{w}dx \int_{-d/2}^{d/2}dy\,\left[xJ_{z}(x,y)\right] \, ,
\end{eqnarray}
will be used. These quantities are straightforwardly calculated from the current distributions $\{ I_i \}$ obtained by the method in Sec.\ref{secCritical}. In practice, $m_{x}$ and $m_{y}$ can be measured by flux detecting coils, conveniently oriented around the sample.

Next, we discuss the influence of $h_{x,a}$ and $h_{y,a}$ on $m_{x}$ and $m_{y}$, so as to provide testing criteria for our theory, and also to allow comparison with the available literature on the subject.
\begin{figure*}[t]
\centerline{
\includegraphics[width=16cm]{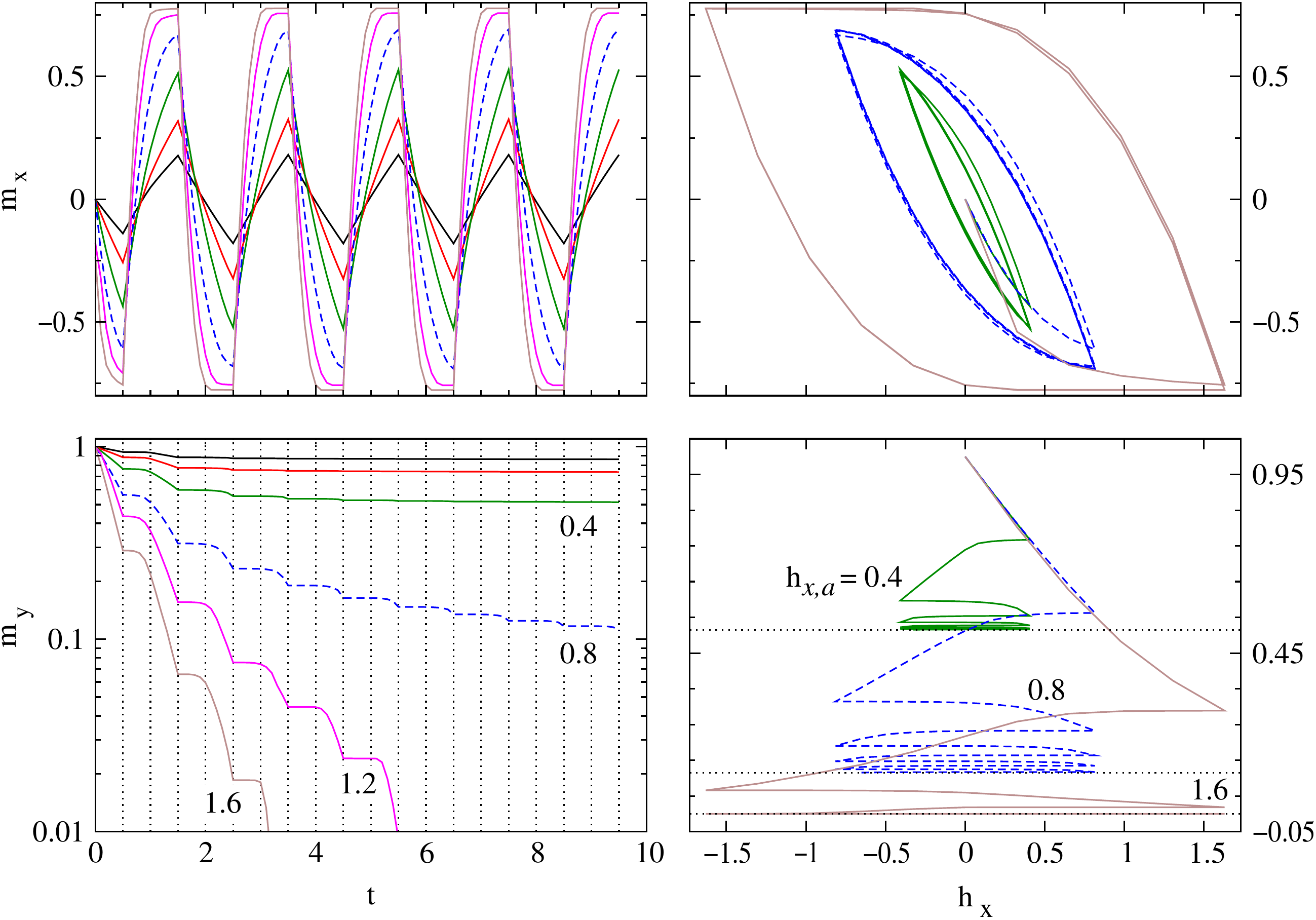}}
\caption{\label{fig7}(Color online) Evolution of the magnetic moment components for several excitation processes in the form sketched in Fig.\ref{fig1}. $m_x$ and $m_y$ are obtained through Eq.(\ref{eqm}) and normalized to the values of $m_y$ at the beginning of the $ac$ cycle. Time ($t$) is given in units of $t_{0}\equiv\tau_{ac}w/2d$. $h_x$ is given in units of the penetration field $H_p$. The different lines (some are labeled for clarity) correspond to $h_{x,a}=0.1,\, 0.2,\, 0.4,\, 0.8,\, 1.2,\, 1.6$. All have been obtained for $h_{y,a}=0.4$}
\end{figure*}

\subsection{Influence of the oscillation amplitude ($h_{x,a}$)}
First, we concentrate on the effect of changing $h_{x,a}$ with $h_{y,a}$ fixed, i.e.: one {\em polarizes} the sample by means of a given value of the vertical field, and performs $ac$ cycles of the horizontal component for different amplitudes. According to our critical state theory, the results of such experiment would be as shown in Fig.\ref{fig7}. Recall that, on taking $h_{y,a}=0.4$ and cycling $h_{x,a}$ between the values $\pm h_{x,a}=\pm 0.1, \pm 0.2, \pm 0.4,\pm 0.8, \pm 1.2, \pm 1.6$, $m_{x}$ and $m_{y}$ display a behavior which is clearly different if one probes either above or below $h_{x,a}=0.8$.

As relates to $m_{x}$, the behavior for low amplitudes is characterized by a non-saturated oscillation that essentially follows the applied field $h_{x}$. When plotted against this quantity $m_{x}$ reaches a stationary Bean's like loop after an initial transient in which the loop does not close. This behavior was already reported in  both experimental\cite{park} and theoretical\cite{badiaprb} works. In the latter, by contrast to the present case, the flux cutting mechanism was considered. The response to higher amplitudes of the oscillating field ($h_{x,a}>0.8$) is characterized by a saturation of $m_x$, which relates to a stationary $m_{x}(h_{x})$ loop, in the typical form of 1D problems, when the full penetration regime is reached.

The properties of $m_y$ are of special mention, since we have found less common features. Firstly, we outline the step-like behavior in the plot $m_{y}(t)$. Note that the descent of this quantity, induced by the oscillation of $h_{x}$, presents a series of {\em plateaus} that are precisely triggered by the return points in $\pm h_{x,a}$. This fact is also apparent within the $m_y(h_{x})$ plot, and was mentioned before in Ref.\onlinecite{vanderbemden}, where a power law $E-J$ model and the {\em high field} region (full penetration) were considered. Recall that here, we also obtain the steps for the low field region. Nevertheless, the early saturation of $m_y$ makes the effect less visible in such cases. 

We want to stress the high similarity of our plot $m_{y}(t)$ and Fig.2 in Ref.\onlinecite{PRLshake}. Thus, the main feature of both graphs is the separation of complete and incomplete relaxation when the value of $h_{x,a}$ approaches one in units of $H_{p}$. On the other side, recall that the  step-like behavior in our plot would be smeared out over a time scale comparable to the one used by the authors in Ref.\onlinecite{PRLshake}. In fact, if one takes the time unit as $t_{0}\equiv \tau_{ac}w/2d$, the approximation for very thin samples used by Brandt and Mikitik leads to $t_{0}\gg \tau_{ac}/2$, while in our case ($w=d$) we have $t_{0}=\tau_{ac}/2$.

\subsection{Influence of the dc field amplitude ($h_{y,a}$)}

\begin{figure}[!]
\centerline{
\includegraphics[width=8.5cm]{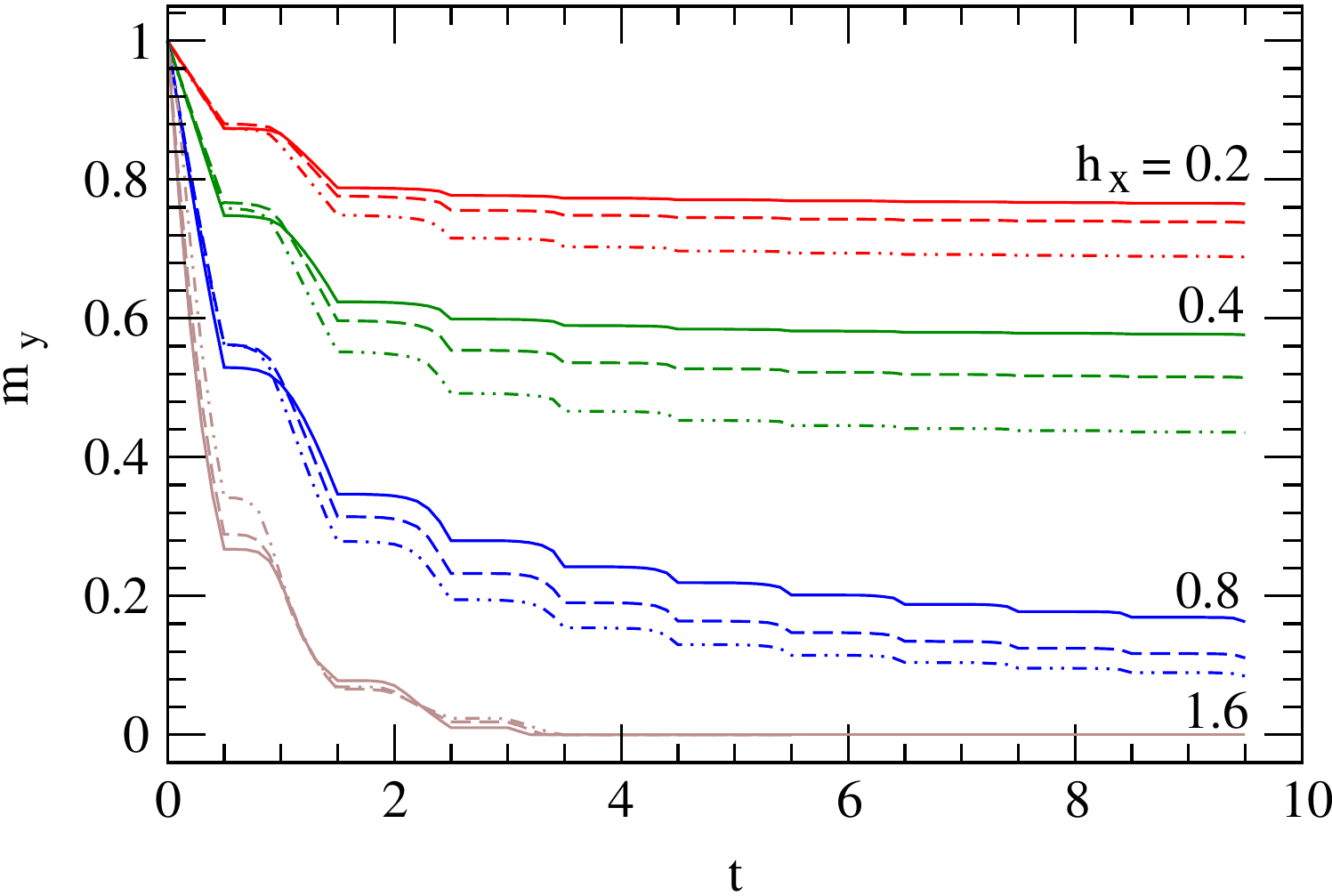}}
\caption{\label{fig8}(Color online) Time dependence of the normalized magnetic moment $m_y$ at various amplitudes of the ac magnetic field $h_{x,a}=0.2,\, 0.4,\, 0.8,\, 1.6$. For each value of $h_{x,a}$ we compare the decay of $m_y$ at three values of $h_{y,a}$: $h_{y,a}=0.2$ (continuous line), $h_{y,a}=0.4$ (dashed line),  $h_{y,a}=0.8$ (dot-dashed)}
\end{figure}

Next, we analyze the relevance of the initial $h_{y}$ ramp on the subsequent complete or incomplete relaxation of $m_{y}$ during the $h_{x}$ cycles. The main results for this study are plotted in Fig.\ref{fig8}. There, we show the evolution of $m_{y}$ for the oscillation amplitudes $h_{x,a}=0.2, 0.4, 0.8, 1.6$ and three different values of the dc field reached in the initial stage of the process: $h_{y,a}=0.2, 0.4, 0.8$. We emphasize two aspects: (i) on the one side, the higher $h_{y,a}$, the lower value of the relaxed component $m_{y,\infty}$ when relaxation is incomplete. However, the separation between the values of $m_{y,\infty}$ first increases, and then decreases, going to zero when the complete relaxation is reached; (ii) on the other side, crossings between the values of the normalized magnetization are observed during the initial transient evolution. The importance of these crossings increases as the value of $h_{y,a}$ does.

\section{Extensions of the theory}
\label{secExtensions}
As an advantage of the mathematical modeling proposed in this work, one can deal with fully arbitrary 2D problems, including non-uniform magnetic sources and non homogeneous properties in the cross section of the sample. The $\{\vec{A},\vec{J}\,\}$-formulation stated in Eqs.(\ref{eqnJJ}) and (\ref{eqMij}) allows to introduce irregular cross sections just by defining the positions of the appropriate elementary wires, and calculating the corresponding $M_{ij}$ matrix.

In this section, we show the modifications introduced by a non homogeneous current carrying capacity. First, we calculate the evolution of the magnetic field profiles for a sample with a central hole, when subjected to the process in Fig.\ref{fig1}. Second, we consider the influence of having a central region with $J_{c}$ noticeably above the corresponding value at the periphery of the sample.

\subsection{Samples with holes}

\begin{figure}[t]
\centerline{
\includegraphics[width=8.5cm]{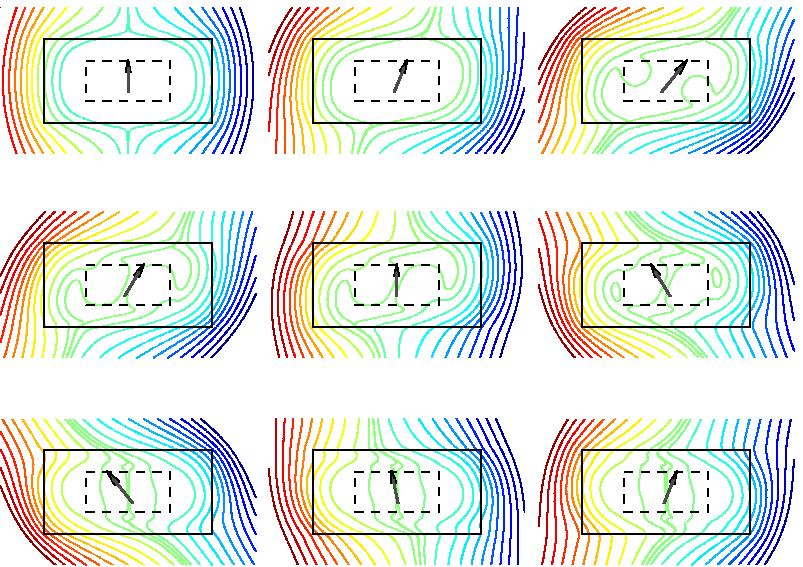}}
\caption{\label{fig9}(Color online) Magnetic flux lines corresponding to the excitation process in Fig.\ref{fig1} for a superconducting bar with a hole (indicated by the dashed line). We show several steps at the transient process within the first cycle.}
\end{figure}

A sample with a hole is straightforwardly treated within the formulation in Eq.(\ref{eqMij}), just by skipping the variables related to the elementary wires within the empty region. Results are shown in Fig.\ref{fig9}. This plot has been obtained for the values of the applied field $h_{x,a}=h_{y,a}=0.4$. Several stages at the first ac cycle are shown, which already suggest the formation of a stationary regime, as in the case of Fig.\ref{fig4}. Nevertheless, in the present situation, the transverse field shaking process cannot induce penetration of current within the empty region. On the other side, the magnetic field lines progressively enter. Then, as a topological effect, the current free and the flux free regions are no longer coincident, subsequent to the contact between the penetrating flux and the hole.

Just for the sake of brevity, we do not show the stationary oscillations which follow after the last stage in Fig.\ref{fig9} (one can easily guess the result). As for the case of homogeneous samples (Fig.\ref{fig4}) they are characterized by a frozen core and swinging flux lines in the peripheral region.

\begin{figure}[b]
\centerline{
\includegraphics[width=8.5cm]{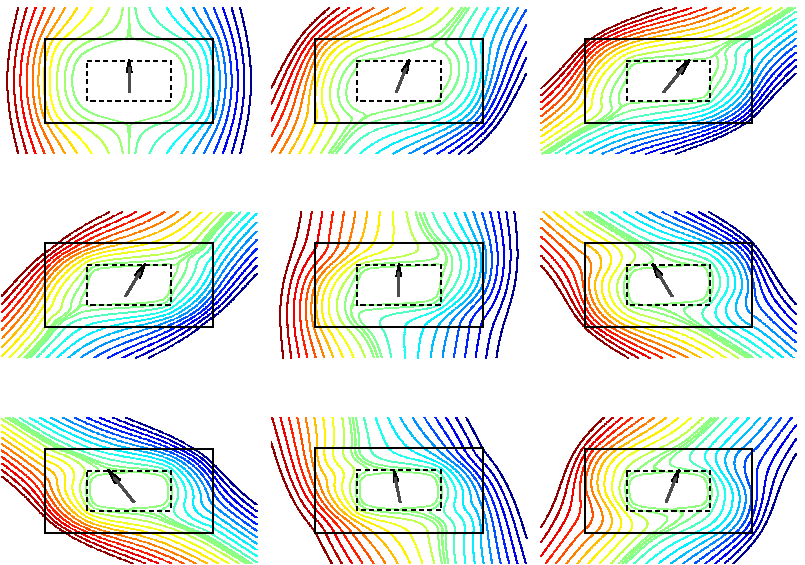}}
\caption{\label{fig10}(Color online) Same as Fig.\ref{fig9} for a full superconducting bar with inhomogeneous pinning properties. $J_c$ within the dashed region is 5 times larger than around.}
\end{figure} 

\subsection{Non homogeneous critical current}

Fig.\ref{fig10} shows the evolution of the flux penetration profiles for a sample with a higher pinning force in the central region. To be specific, we have used the ratio $J_{c,in}=5 J_{c,around}$ in the simulation, i.e. the constraint for the corresponding elements is $-5I_{c}\leq I_{i}\leq 5I_{c}$. In dimensionless units, the applied field amplitudes are again $h_{x,a}=h_{y,a}=0.4$.

As one could expect, both flux and current penetration into the central region are hindered by the higher pinning force. Note that when the initial flux free core is distorted by the oscillation of $h_x$, its boundary slips along the separation between the two regions. On the other hand, the establishment of the stationary regime basically consists of a {\em vertical} critical state profile within the high $J_{c}$ region, and the surrounding swing of flux lines described in the previous examples.

\section{Concluding remarks}
\begin{figure}[b]
\centerline{
\includegraphics[width=8.5cm]{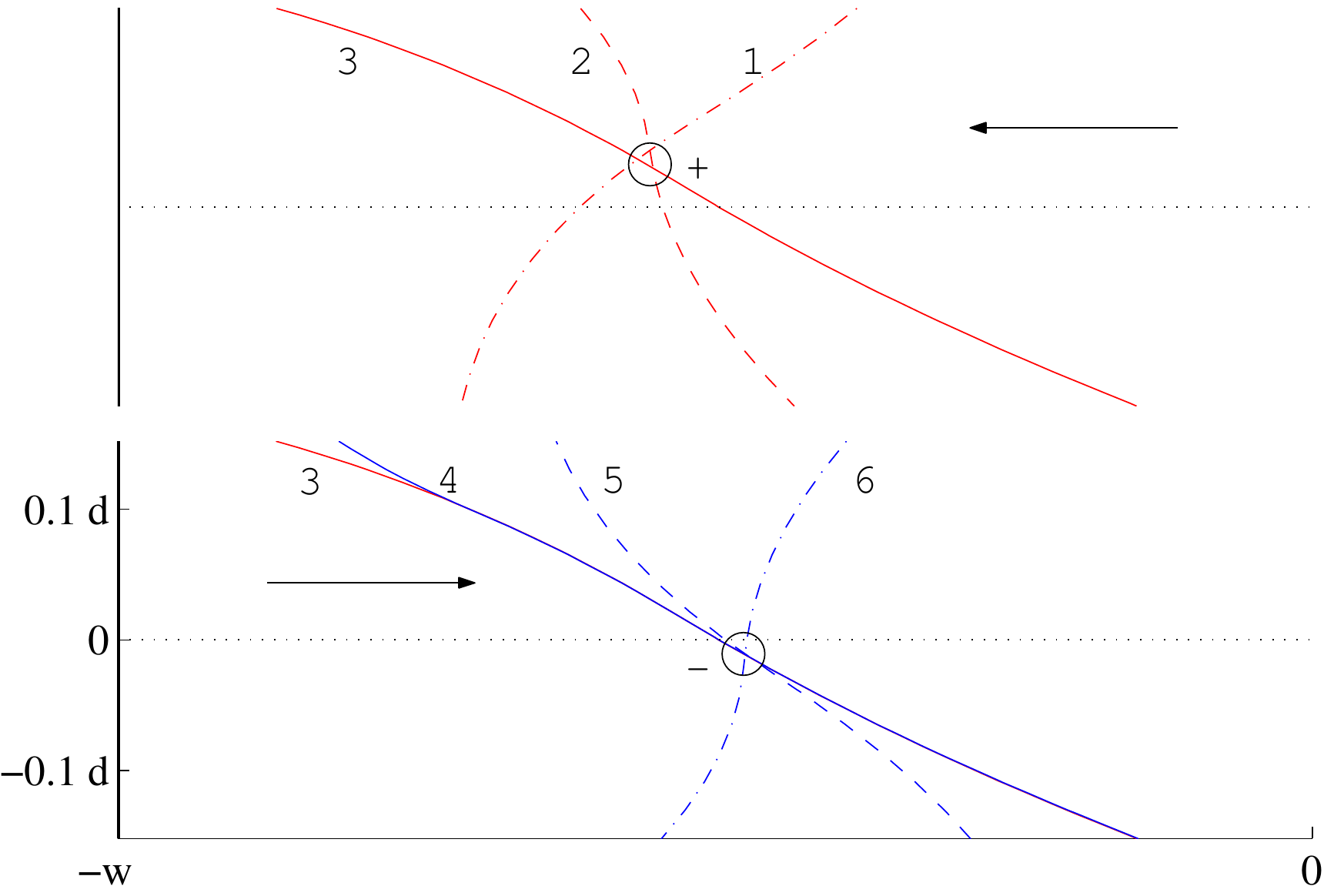}}
\caption{\label{fig11}A flux line moves to the right, by {\em pivoting} around the positions marked (+) and (-), when the ac field undergoes a cycle (sequence of numbers $1-6$). The flux line has been selected from the set of profiles obtained after the conditions in Fig.\ref{fig6}. The picture is split in two panels for clarity.}
\end{figure} 
Although the majority of theoretical and experimental studies about flux pinning in superconductors have focused on the dynamics induced by single-component applied magnetic fields, increasing interest in multi-component situations is arising. Along this line, some recent papers\cite{PRLshake,vanderbemden,clem,badiaprb,park,fisher} have considered the importance of transverse flux effects. Of particular interest is the magnetization decay induced by the oscillations of a perpendicular magnetic field. In this work, we have tried to clarify some questions about the ability of phenomenological flux pinning models for describing the experimental observations. In particular, we have addressed the debate about the existence of complete and incomplete relaxation regimes (full decay or not into the equilibrium magnetization) in terms of the actual transverse shaking process. We have shown that, within the critical state model for flux pinning, the kind of relaxation is related to the existence of a flux free core within the sample. Thus, under certain conditions (small enough amplitudes of the applied field components) one finds a transient regime in which the central flux free core changes its shape until a stationary profile is reached. Afterwards, the transverse field shaking produces the swing of the flux lines around the frozen core structure, and the magnetic moment becomes constant (but not zero). If the amplitude of the applied magnetic field increases enough, the flux free core shrinks to null volume, and the stationary  magnetic moment becomes zero. 

The existence of incomplete/complete relaxation can be observed in experimental works\cite{fisher} and is also predicted by the vortex-shaking model in Ref.\onlinecite{PRLshake}. Although there are some differences between the statement of that model and ours, the essential implications are easily reconciled. To start with, the model in Ref.\onlinecite{PRLshake} is a good approximation for thin strips, which are nothing but a limiting case of the 2D problem. On the other hand, it is of mention that a fundamental ingredient of Ref.\onlinecite{PRLshake} is the idea that flux lines {\em walk} towards the center of the sample ($x=0$), pivoting around the so-called {\em swivel points}. Switching from one swivel point to the next is done every half-cycle. In the present work, we have merely used the concept of a maximum pinning force (introduced by the phenomenological parameter $J_{c\bot}$, through the restriction $|J_{\bot}|\leq J_{c\bot}$ in the variational solution of Maxwell equations). Remarkably, the evolution of flux lines in our model follows the walking pattern introduced by Brandt and Mikitik. This is shown in Fig.\ref{fig11}. Starting from the magnetic field profiles related to the process in Fig.\ref{fig6}, we have plotted a sequence of pictures, corresponding to a specific flux line, along the first cycle. It is apparent that two swivel points are formed (labeled $+$ and $-$), which establish the shift from position $1$ to $6$, by two rotations. The switch from the pivot $+$ to $-$ is clearly established by the return point $-H_{x,a}$ (step $3$ to $4$). We want to note that the vertical position of the swivel points has changed.

Another point to consider is the role of the aspect ratio parameter $w/d$. Thus, the main difference between the predicted magnetization decay, that is, the step-like structure in our Figs.\ref{fig7} and \ref{fig8} disappears when the physical time unit $t_{0}\equiv\tau_{ac}w/2d$ is taken into account. The steps are smeared out by the time scale of the plot, and the two models coincide again.

Finally, a comparison between our proposal with the method introduced in Ref.\onlinecite{vanderbemden} ($E\propto(J/J_{c})^{n}$) also seems to be of interest. In principle, the results of this article should be obtained as the limit of the former, when higher and higher values of $n$ are used. This would allow to quantify the importance of flux creep effects, intrinsically included in phenomenological $E-J$ models.

\section*{Acknowledgments}
The authors acknowledge financial support from Spanish CICYT
(Project Nos. BMF-2003-02532 and MAT2005-06279-C03-01).

\end{document}